\documentclass{article}
\begin{document}

\begin{center}
\centerline{\large \bf Femtosecond laser as a tool for experimental study }
\centerline{\large \bf of inequality of forward and reversed processes in 
optics.}
\end{center}

\vspace{3 pt}
\centerline{\sl V.A.Kuz'menko\footnote{Electronic 
address: kuzmenko@triniti.ru}}

\vspace{5 pt}
\centerline{\small \it Troitsk Institute for Innovation and Fusion 
Research,}
\centerline{\small \it Troitsk, Moscow region, 142190, Russian 
Federation.}
\vspace{5 pt}
\begin{abstract}

The classical pump-probe scheme for experiments with femtosecond laser is 
proposed to use for study the inequality of  forward and reversed transitions 
in optics.

\vspace{5 pt}
\end{abstract}

\vspace{12 pt}

	Rather usual situation in nonlinear optics is a case, when the 
mathematical description of phenomenon is not accompanied by corresponding 
clear explanation of physical nature of this phenomenon. The most significant 
example is the effect of adiabatic population transfer in a two level system 
after sweeping of resonance conditions. The mathematical model of rotating 
wave approximation gives excellent description of phenomenon, but it physical 
explanation is absent [1,2].

	Similar situation exists in the field of rotational coherence 
spectroscopy [3]. Corresponding mathematical model well describes the 
temporal characteristics and the shape of observed rotational transients 
(revivals) [4]. As a physical explanation of the nature of transients the 
concept of quantum interference of coherent states is usually used [3]. 
However, the opinion exists that the concept of coherent states has not 
reliable physical base [5-7]. Good alternative here is the concept of 
inequality of forward and reversed processes in optics [8]. Now we have not 
only indirect evidences, but also the quite direct experimental proofs of 
such inequality [9-11]. 

	The experiments show that the reversed process can have extremely 
high efficiency. For optical transitions it corresponds to very high 
cross-section of reversed transition into the initial state. However, the 
integral cross-sections of forward and reversed transitions (Einstein's 
coefficients) should be equal. Some indirect evidences allow to suppose the 
following shape of orientation dependence of cross-section for reversed 
transition: extremely high cross-section for transition into the initial 
state and relatively small cross-section for other orientation of molecules 
in space (Fig.1). In this case the explanation of physical origin of 
rotational coherence and population transfer effects become quite simple 
and clear. 

        The main way for experimental study of transition cross-section 
dependence from orientation of molecule relative to the direction of laser 
beam is quite clear: we can use a sequential optical transitions [12]. 
The radiation of the first laser selects the molecules from isotropic 
distribution, the second laser beam deals with the aligned molecules and 
the third laser beam will deals with the oriented molecules. However, any 
complete and extensive experimental works of such kind are absent in 
literature. 
       
	Some negative role here, obviously, plays the concept of so-called 
pendular states [12].  It is supposed, that intense laser radiation can 
break a free rotation of molecules and align it about the laser polarization 
axis. Extensive experimental works in this field, however, can not give any 
reliable proofs, that this effect really exists [12]. The alternative 
explanation of the experimental results, which includes the concept of 
inequality of forward and reversed transitions, is quite possible [13]. 

	Pump-probe experiments with femtosecond laser may be used for study 
the orientation spectra of both forward and reversed transitions. Large 
number of pump-probe experiments in the field of rotational coherence 
spectroscopy are described in literature. But in most cases the energy of 
the probe pulse is equal to the energy of the pump pulse [4,14,15] or even 
exceeds it [16]. The scientists usually take an interest in temporal 
characteristics and in shape of revivals.

	The main purpose of this note is to propose to use for experimental 
study of cross-sections of forward and reversed transitions the classical 
pump-probe scheme, when the energy of probe pulse is much smaller, than the 
energy of pump pulse. In this case the radiation of probe pulse does not 
change substantially the distribution of molecules after the pump pulse and 
does not make difficulties for interpretation of experimental results. 
Measured amplitude of transients contains information about the 
cross-section of forward and reversed transitions.

	Two different proposed experimental schemes are presented in Fig.2. 
In the first case (Fig.2a, where the excited state takes part) the dependence 
of integral fluorescence intensity of molecules is studied from the delay 
time between pump and probe pulses. The energy of the probe pulse should 
be 0,01 - 0,05 of the pump pulse. It is advisable to modulate intensity of 
probe beam and to use a phase-sensitive detection scheme for precise 
measurement of absorption or amplification of probe pulse radiation. We can 
expect to observe very strong transients with even some amplification of 
probe pulse (similar in shape as in Fig.5 of Ref. [9]). If we know the 
fraction of molecules excited by the pump pulse, we can determine the 
ratio of cross-section of forward and reversed transitions.

	The second case is very similar to work [4] (Fig.2b, Raman variant 
without an excited states of molecules), where the modification of 
polarization of reading pulse is measured. This reading pulse is tuned to 
one of the far transients. The probe pulse has the same polarization as the 
pump pulse. In this case it is also helpful to use the phase-sensitive 
detection scheme with modulated probe beam. We can expect, that the 
scanning of the probe pulse will give some decreasing of measured signal 
due to efficient transfer of molecules into the initial state. The 
dependence of signal from intensity of pump and probe pulses will give 
information about the cross-section of forward and reversed transitions. 

	In conclusion, we expect that the discussed experiments will make 
clear a physical origin of large number of phenomena in nonlinear optics.

\end{document}